\documentclass[
    ,final            
  ]
  {aipproc}

\usepackage{amsmath,amssymb}
\usepackage{bm}

\layoutstyle{8x11double}

\begin{document}

\title{Quantum Critical Point of Itinerant Antiferromagnet in 
the Heavy Fermion Ce(Ru$_{1-x}$Rh$_{x}$)$_{2}$Si$_{2}$}


\author{Hiroaki Kadowaki}{
  address={Department of Physics, Tokyo Metropolitan University, 
Hachioji-shi, Tokyo 192-0397, Japan}
}

\author{Yoshikazu Tabata}{
  address={Graduate School of Science, Osaka University, 
Toyonaka, Osaka 560-0043, Japan}
}

\author{Masugu Sato}{
  address={MSD, JASRI, 
1-1-1 Kouto Mikazuki-cho Sayo-gun, Hyogo 679-5198, Japan}
}

\author{Naofumi Aso}{
  address={NSL, Institute for Solid State Physics, 
University of Tokyo, Tokai, Ibaraki 319-1106, Japan}
}

\author{Stephane Raymond}{
  address={CEA-Grenoble, DSM/DRFMC/SPSMS, 38054 Grenoble, France}
}

\author{Shuzo Kawarazaki}{
  address={Graduate School of Science, Osaka University, 
Toyonaka, Osaka 560-0043, Japan}
}

\begin{abstract}
A focus of recent experimental and theoretical studies on heavy 
fermion systems close to antiferromagnetic (AFM) quantum critical 
points (QCP) is directed toward revealing the nature of the 
fixed point, i.e., whether it is an itinerant antiferromagnet 
[spin density wave (SDW)] type or a locally-critical fixed point. 
The relevance of the local QCP was proposed to explain the $E/T$-scaling 
with an anomalous exponent observed for the AFM QCP of 
CeCu$_{5.9}$Au$_{0.1}$. 
In this work, we have investigated an AFM QCP of another archetypal 
heavy fermion system Ce(Ru$_{1-x}$Rh$_{x}$)$_{2}$Si$_{2}$ with 
$x = 0$ and 0.03 ($\sim x_{\textrm{c}}$) using single-crystalline neutron 
scattering. 
Accurate measurements of the dynamical susceptibility 
Im$\chi (\bm{Q},E)$ at the AFM wave vector 
$\bm{Q} = 0.35\bm{c}^{\ast}$ have shown that Im$\chi (\bm{Q},E)$ 
is well described by a Lorentzian and its energy width $\Gamma (\bm{Q})$, 
i.e., the inverse correlation time 
depends on temperature as 
$\Gamma (\bm{Q}) = c_{1} + c_{2} T^{3/2 \pm 0.1}$, 
where $c_1$ and $c_2$ are $x$ dependent constants, in low 
temperature ranges.
This critical exponent $3/2$ 
proves that the QCP is controlled by the SDW QCP 
in three space dimensions studied by the 
renormalization group and self-consistent renormalization theories. 
\end{abstract}

\maketitle


\section{Introduction}

\hspace*{2.5 mm} Quantum critical points (QCP) separating ferromagnetic or 
antiferromagnetic states from paramagnetic Fermi liquid 
states in strongly correlated electron systems have been 
investigated for decades. 
Successful descriptions of quantum critical behavior 
were provided by the self-consistent renormalization 
(SCR) theory of spin fluctuations \cite{MoriyaBook85,Moriya-Ueda03} 
for $d$-electron systems based on the Hubbard model. 
The mean-field type approximations in this theory were justified 
by the renormalization group studies \cite{Hertz76,Millis93} 
using Hertz's effective action above upper critical dimensions. 
For the ferromagnetic QCP, theoretical predictions were 
supported by experimental studies of $d$-electron systems 
\cite{MoriyaBook85}. 
In contrast there is little experimental understanding 
of the antiferromagnetic QCP \cite{Moriya-Ueda03}. 

A recent intriguing issue of QCP under controversial 
debate is directed toward revealing relevant fixed points for 
antiferromagnetic QCPs in heavy-fermion systems \cite{Coleman01}. 
For energy scales much lower than the Kondo temperature $T_{\text{K}}$, 
$f$ and conduction electrons form composite 
quasiparticles with a large mass renormalization in 
paramagnetic heavy fermions. 
By tuning a certain parameter, e.g., pressure or concentration, 
an antiferromagnetic long range order emerges 
from the Fermi liquid state. 
In a weak coupling picture, it has been hypothesized that 
the same QCP as the $d$-electron itinerant antiferromagnet, 
referred to as spin density wave (SDW) type QCP, is relevant 
to the heavy fermion QCP \cite{Moriya-Taki,Coleman01}. 

However despite a number of experimental studies 
of heavy-fermion systems showing non-Fermi liquid behavior, 
none of them definitively supports this QCP 
\cite{Lohneysen94,Kambe96,raymond01a,Stewart01}. 
This stems partly from experimental difficulty in measuring 
weakly divergent quantities around a QCP, 
especially for bulk properties, 
which has been also the case 
for $d$-electron itinerant antiferromagnets \cite{Moriya-Ueda03}. 
On the contrary, several recent experiments suggest 
the possibility of a novel strong coupling picture of the 
QCP \cite{Coleman01,Schroder00,Paschen04}.
Among these studies, 
single-crystalline neutron scattering investigations of 
the heavy fermion CeCu$_{5.9}$Au$_{0.1}$ 
provided interesting insight \cite{Schroder00}. 
On the basis of the observed 
$E/T$-scaling with an anomalous exponent \cite{Schroder00}, 
and effective two space dimensions \cite{Stockert98}, 
a scenario of a locally critical 
QCP was proposed \cite{Coleman01,Si01}. 

In this work, we have studied the antiferromagnetic 
QCP of another heavy-fermion 
system Ce(Ru$_{1-x}$Rh$_{x}$)$_2$Si$_2$ ($x=0$, $0.03$) 
using single-crystalline neutron scattering 
\cite{Kadowaki05}. 
Stoichiometric CeRu$_2$Si$_2$ is an archetypal paramagnetic 
heavy-fermion with enhanced $C/T \simeq 350$ mJ/K$^2$ mol 
and $T_{\text{K}} \simeq 24$ K \cite{Besnus85}. 
Extensive neutron scattering studies of 
CeRu$_2$Si$_2$ \cite{Kadowaki04} 
have shown that spin fluctuations possessing 
three-dimensional ($d=3$) character are excellently 
described by the SCR theory for heavy fermions \cite{Moriya-Taki}. 
A small amount of Rh doping, 
$x > x_{\text{c}} \simeq 0.04$ \cite{Sekine92}, 
induces an antiferromagnetic phase 
(see the inset of Fig.~\ref{fig:Escan})
of the sinusoidally modulated structure 
with the wave vector $\bm{k}_3 = 0.35 \bm{c}^{\ast}$ 
\cite{Kawarazaki97}. 
Samples nearly tuned to the lowest concentration 
QCP ($x \sim x_{\text{c}}$) show non-divergent 
$C/T$ ($T \rightarrow 0$) \cite{Tabata98} and 
$\Delta \rho \propto T^{3/2}$ \cite{Kadowaki05}, 
which are consistent with the SDW QCP in $d=3$. 
Thus one can expect that Ce(Ru$_{1-x}$Rh$_{x}$)$_2$Si$_2$ 
($x \lesssim x_{\text{c}}$) is suited to 
investigate the SDW QCP without disorder effects. 

\section{Experiments and analyses}

\hspace*{2.5 mm} Neutron-scattering measurements were performed on the triple-axis 
spectrometer HER at JAERI. 
It was operated using final energies of 
$E_{\text{f}}=3.1$ and $2.4$ meV providing energy resolutions 
of $0.1$ and $0.05$ meV 
(full width at half maximum), respectively, at elastic positions. 
Single crystals with a total weight of 19 g ($x=0$) and 
17 g ($x=0.03$) were grown by the 
Czochralski method. 
Two sets of multi-crystal samples aligned together 
were mounted in a He flow cryostat so as to measure 
a $(h0l)$ scattering plane. 
All the data shown are converted to the dynamical susceptibility 
$\text{Im}\chi(\bm{q}, E)$. 
It is scaled to absolute units by comparison with the intensity of 
a standard vanadium sample. 
We note that a new point of the present work is unprecedented 
experimental accuracy in determining the critical 
exponent using large samples and long counting time. 
This has enabled us to determine the singularity 
of the QCP and to make qualitative conclusions of the universality class, 
which was very difficult in the pioneering work using 
the related compound Ce$_{1-x}$La$_{x}$Ru$_2$Si$_2$ 
\cite{Kambe96,raymond01a}. 

\subsection{Previous Study of CeRu$_2$Si$_2$}

\begin{figure}
\includegraphics[width=7.8cm,clip]{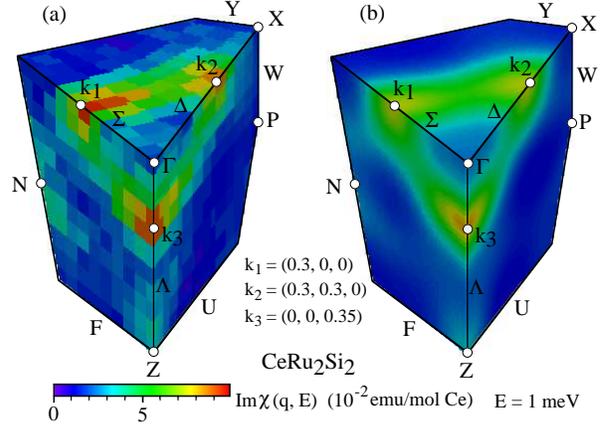}
\caption{
\label{fig:3Dmap}
Observed (a) and calculated (b) 
intensity maps of constant-$E$ scans taken with $E = 1$ meV 
at $T=1.5$ K for the sample with $x = 0$ \cite{Kadowaki04}. 
They are shown on the surface of the irreducible Brillouin zone. 
}
\end{figure}

Let us first make brief comments on our previous 
neutron scattering study 
of CeRu$_2$Si$_2$ \cite{Kadowaki04}. 
We showed that spin fluctuations of CeRu$_2$Si$_2$ are 
reasonably well described by the SCR theory 
for heavy fermions \cite{Moriya-Taki}. 
This result has the following two implications in connection with the SDW QCP. 
First, the spin fluctuations at low temperature $T = 1.5$ K 
can be parametrized by the SCR form \cite{Moriya-Taki} 
\begin{equation}
\label{eq:DS}
\chi (\bm{q}, E)^{-1} 
= 
\chi_{\text{L}} (E)^{-1} 
- J(\bm{q})
\; .
\end{equation}
This equation means that the local dynamical susceptibility 
$\chi_{\text{L}} (E) = 
\chi_{\text{L}}/(1 - \text{i} E/ \Gamma_{\text{L}})$, 
expressing the local quantum fluctuation by the Kondo effect, 
is modulated by the intersite exchange interactions 
$J_{\bm{r}, \bm{r}'}$ [$J(\bm{q}) = 
\sum_{\bm{r} \neq 0} J_{\bm{r},0} \exp( i \bm{q} \cdot \bm{r})$]. 
A number of constant-$Q$ and -$E$ scan spectra can be reproduced by 
Eq.~(\ref{eq:DS}) with 
the adjustable parameters 
of $\chi_{\text{L}}$, $\Gamma_{\text{L}}$, and 
14 exchange parameters \cite{Kadowaki04}. 
In Fig.~\ref{fig:3Dmap} we show observed and calculated intensity 
maps of constant-$E$ scans with $E = 1$ meV at $T=1.5$ K. 
One can see that the calculated intensity reproduces 
the complicated antiferromagnetic spin fluctuations 
with the three peaks and weaker structures around the Z and N points.

Second, the temperature dependence of the spin fluctuations 
of CeRu$_2$Si$_2$ \cite{Kadowaki04} 
can be approximately described by the SCR 
scenario \cite{Moriya-Taki}, in which 
the $T$ dependence of Eq.~(\ref{eq:DS}) is 
brought about by the single $T$ dependent parameter 
$\chi_{\text{L}}(T)$. 
The temperature dependence of $\chi_{\text{L}}(T)$ is determined 
by the self-consistent relation \cite{Moriya-Taki}. 
The existence of this single $T$ dependent parameter indicates 
that the underlying mechanism is controlled by a neighboring 
SDW QCP. 
In fact, the numerically calculated $\chi_{\text{L}}(T)$, 
showing $\chi_{\text{L}}(T) \propto \text{const} - T^{3/2}$ 
\cite{Kadowaki04} in a low temperature range, agrees with the  
computation [Eq.~(\ref{eq:T1.5})] by 
the renormalization group theory of the SDW QCP \cite{SachdevBook99}. 
Along this line, the purpose of this work is to check whether the 
$T^{3/2}$ dependence of $\chi_{\text{L}}(T)$ really occurs in 
CeRu$_2$Si$_2$ and 
in the nearly tuned sample 
Ce(Ru$_{0.97}$Rh$_{0.03}$)$_2$Si$_2$.

\subsection{QCP of Ce(Ru$_{1-x}$Rh$_{x}$)$_2$Si$_2$}

In the scenario of the SDW QCP in $d=3$, which was well established 
by the renormalization group theory 
\cite{Hertz76,Millis93,SachdevBook99}, 
the wave-vector dependent susceptibility for the tuned 
sample ($x = x_{\text{c}}$) diverges 
as $\chi(\bm{k}_3) \propto T^{-3/2}$ 
\cite{MoriyaBook85,Millis93}, 
or the characteristic energy of spin fluctuation, 
i.e., the inverse correlation time, 
depends on temperature as 
$\Gamma(\bm{k}_3) \propto \chi(\bm{k}_3)^{-1} \propto T^{3/2}$. 
By taking the detuning effect ($x < x_{\text{c}}$) into account, 
the leading two terms of $\Gamma(\bm{k}_3)$ 
computed by the renormalization group theory 
\cite{Millis93,SachdevBook99} 
are given by 
\begin{equation}
\label{eq:T1.5}
\Gamma(\bm{k}_3) 
=
 c_1 + c_2 T^{3/2} 
\; ,
\end{equation}
where $c_{1}$ and $c_{2}$ are $x$ dependent constants. 
This equation is an approximation in the temperature range 
$T_{\text{FL}} \ll T \ll T_{\text{K}}$, 
where $T_{\text{FL}}$ is a crossover temperature below which 
the system shows the Fermi liquid behavior 
\cite{Millis93,SachdevBook99}. 

\begin{figure}
\includegraphics[width=7.1cm,clip]{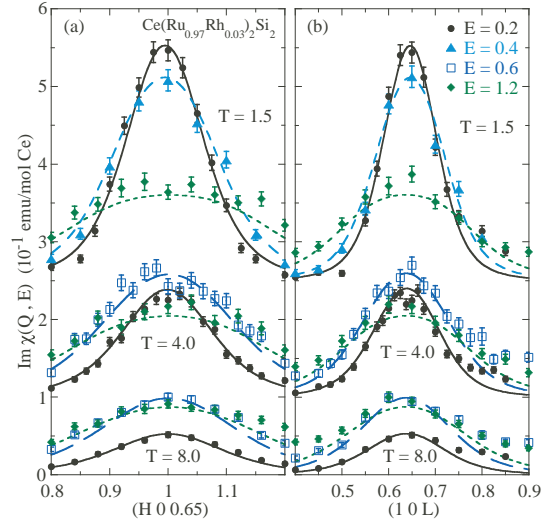}
\caption{
\label{fig:Qscan}
Constant-$E$ scans taken with 
$E = 0.2$, 0.4, 0.6 and 1.2 meV 
along the $\bm{Q} = (H,0,0.65)$ (a) and $(1,0,L)$ (b) lines 
for the sample with $x = 0.03$. 
Data at $T = 1.5$ and 4 K are shifted by 
0.25 and 0.1 emu/mol Ce, respectively for clarity.
The curves are calculations using Eqs.~(\ref{eq:ImChi}), 
corrected for resolution functions with the same fit parameters as 
those shown in Fig.~\ref{fig:Escan}. 
}
\end{figure}

The imaginary part of the dynamical susceptibility 
at $\bm{Q}=\bm{k}_{3}+\bm{q}$ 
with small $|\bm{q}|$ and $|E|$
is approximated \cite{SachdevBook99} by 
\begin{subequations}
\label{eq:ImChi}
\begin{eqnarray}
\text{Im}\chi(\bm{k}_{3}+\bm{q}, E) 
&=&
\frac{ \chi(\bm{k}_{3}) \Gamma(\bm{k}_{3}) E }
{ E^2 + \Gamma(\bm{k}_{3}+\bm{q})^2 }
\; ,\label{eq:ImChi1}
\\
\Gamma(\bm{k}_{3}+\bm{q}) 
&=& 
D_c [\kappa_c^2 + q_c^2 + F (q_a^2 + q_b^2)]
\, ,\label{eq:ImChi2}
\end{eqnarray}
\end{subequations}
[a quadratic expansion of Eq.~(\ref{eq:DS})] 
where $D_c$ and $F$ are $T$ independent parameters, and 
$\kappa_c$ is the inverse correlation length along the $c$-axis. 
In Eq.~(\ref{eq:ImChi1}) the product $\chi (\bm{k}_{3}) \Gamma (\bm{k}_{3})$ 
is theoretically $T$ independent. 
The two parameters $D_c$ and $F$ were determined by 
using constant-$Q$ and \mbox{-$E$} scans for both samples with 
$x=0$ \cite{Kadowaki04} and $0.03$ at $T=1.5$ K. 
These data were fit to Eqs.~(\ref{eq:ImChi}) 
convolved with the resolution functions. 
In Fig.~\ref{fig:Qscan} we show constant-$E$ scans 
through the antiferromagnetic wave vector $\bm{Q}=(101)-\bm{k}_{3}$, 
and the fit curves for the sample with $x=0.03$. 
The good quality of fitting indicates that 
Eqs.~(\ref{eq:ImChi}) well describe the experimental data 
at $T=1.5$ K. 
We obtained $x$ independent values of the 
parameters $D_c = 98 \pm 4$ (meV \AA$^2$) and 
$F = 0.12 \pm 0.01$. 

\begin{figure}
\includegraphics[width=7.1cm,clip]{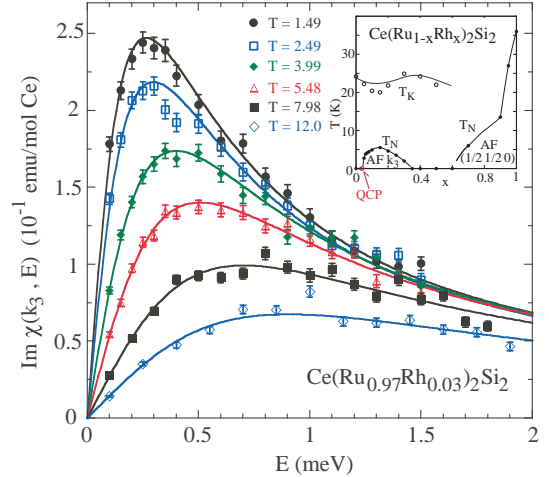}
\caption{
\label{fig:Escan}
Constant-$Q$ scans 
measured at the antiferromagnetic wave vector 
$\bm{Q}=(101)-\bm{k}_{3}$ 
for the sample with $x=0.03$.
The curves are fit results using Eq.~(\ref{eq:ImChi1}) with 
two adjustable parameters, 
$\Gamma (\bm{k}_{3})$ and $\chi (\bm{k}_{3})$. 
The inset shows the phase diagram and $T_{\text{K}}$ 
of Ce(Ru$_{1-x}$Rh$_{x}$)$_2$Si$_2$~\cite{Sekine92,Kawarazaki97}. 
}
\end{figure}

The temperature dependence of 
$\Gamma (\bm{k}_{3}) = D_{c} \kappa_{c}^{2}$ for 
both samples has been determined 
by performing constant-$Q$ scans taken at 
$\bm{Q}=(101)-\bm{k}_{3}$. 
These scan data were fit to Eqs.~(\ref{eq:ImChi}) 
convolved with the resolution functions, 
where there are two adjustable parameters 
$\Gamma (\bm{k}_{3})$ and $\chi (\bm{k}_{3})$. 
Several fit results of the constant-$Q$ scans 
for $x=0.03$ are shown in Fig.~\ref{fig:Escan}, 
where one can see that the quality of fitting 
is excellent. 
We also checked the $T$ independence of the 
parameters $D_c$ and $F$ 
by comparing the constant-$E$ scans in Fig.~\ref{fig:Qscan} 
at $T=4$ and 8 K with those calculated using the $T$ dependent 
$\Gamma (\bm{k}_{3})$ and $\chi (\bm{k}_{3})$ determined 
by the constant-$Q$ scans. 
The calculated curves in Fig.~\ref{fig:Qscan} agree reasonably 
well with the observations. 
Thus we conclude that the theoretical approximation of 
Eqs.~(\ref{eq:ImChi}) has been experimentally confirmed, 
and that the fit parameter $\Gamma (\bm{k}_{3})$ has been 
determined very precisely. 

\begin{figure}
\includegraphics[width=7.1cm,clip]{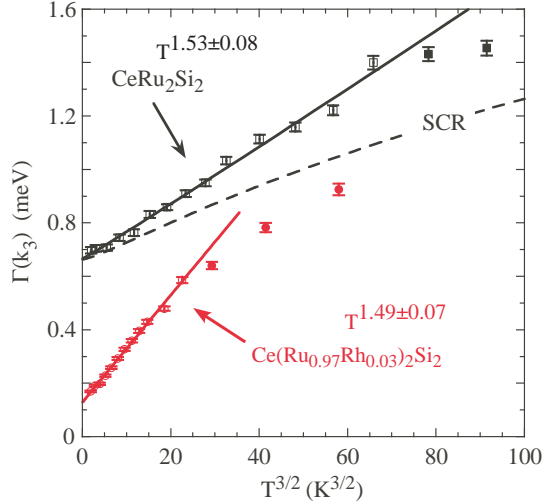}
\caption{
\label{fig:Gammak3}
Energy width $\Gamma (\bm{k}_{3})$ 
of the Lorentzian form [Eq.~(\ref{eq:ImChi1})] 
is plotted as a function of $T^{3/2}$. 
The full lines are fits to $\Gamma (\bm{k}_{3}) = c_1 + c_2 T^{v}$ 
with adjustable parameters $c_1$, $c_2$, and $v$ 
in the low temperature ranges. 
The dashed line is the calculation \cite{Kadowaki04} 
using the SCR theory \cite{Moriya-Taki}. 
}
\end{figure}

The temperature dependence of $\Gamma (\bm{k}_{3})$ is 
shown in Fig.~\ref{fig:Gammak3} by plotting data 
as a function of $T^{3/2}$. 
At low temperatures the observed data clearly agree with 
the linear behavior of Eq.~(\ref{eq:T1.5}). 
In fact, by least squares fitting we obtained: 
$\Gamma (\bm{k}_{3}) = (0.67 \pm 0.01) 
+ (0.0095 \pm 0.0021) T^{1.53 \pm 0.08}$ 
(in units of meV) in the range 
$1.5<T<16$ K for the sample with $x=0$, and 
$\Gamma (\bm{k}_{3}) = (0.129 \pm 0.007) 
+ (0.020 \pm 0.003) T^{1.49 \pm 0.07}$ 
in the range $1.5<T<8$ K for $x=0.03$. 
Therefore we conclude that the observed critical 
exponent of $3/2 \pm 0.1$ is in agreement with the 
theoretical value of $3/2$ and, consequently, that the 
temperature dependence of spin fluctuation 
is controlled by the SDW QCP in $d=3$. 
The same exponent for both $x=0$ and 0.03 samples 
ensures that randomness due to Rh doping does not 
affect the criticality.

\section{Discussion and Conclusion}

\hspace*{2.5 mm} The constant $c_1$ in Eq.~(\ref{eq:T1.5}) is proportional to 
the theoretical tuning parameter 
and, hence, $c_{1} \propto x_{\text{c}}-x$ is normally assumed 
\cite{Millis93}. 
This assumption is consistent with the observed values of 
$c_1$ and the critical concentration 
$x_{\text{c}}=0.04 \pm 0.005$. 
On the other hand, the constant $c_2$ 
is theoretically assumed to be weakly dependent on $x$ \cite{Millis93}. 
However we observed appreciable $x$ dependence 
$c_2(x=0.03)/c_2(x=0) \sim 2$, which may suggest certain 
unknown perturbations. 
Despite this problem, we think that the critical exponent of $3/2$, 
which is determined solely by basic characteristics of the 
system (the space dimension $d=3$ and the dynamical 
exponent $z=2$), 
is more important and decisive to conclude 
the nature of the QCP.

An advantage of the present neutron scattering study is that 
Eq.~(\ref{eq:T1.5}) holds in a wider temperature range 
compared to those of indirect measurements of bulk properties. 
For example, the leading terms of the specific heat 
$C/T = \gamma_0 - \alpha T^{1/2}$ were shown to have too narrow 
$T$ ranges to be clearly observed \cite{Kambe96,Moriya-Taki}. 
In Fig.~\ref{fig:Gammak3}, the dashed line reproduces 
the SCR computation of $\Gamma (\bm{k}_{3})$ 
for CeRu$_2$Si$_2$ \cite{Kadowaki04}. 
Apart from discrepancy of the coefficient $c_2$, one can see that 
the $T^{3/2}$ dependence of Eq.~(\ref{eq:T1.5}) 
is a good approximation for the SCR curve in the 
$T$ range $2.5 < T < 13$ K ($4<T^{3/2}<47$). 

In connection with the neutron scattering study of 
CeCu$_{5.9}$Au$_{0.1}$ \cite{Schroder00}, 
it was theoretically predicted ~\cite{Si01} 
that the locally critical QCP is relevant for the 
two-dimensional spin fluctuation \cite{Stockert98}. 
This theory also predicted that the SDW QCP is relevant for 
the three-dimensional spin fluctuation, being 
in accord with the present results. 
Finally we note that, to our knowledge, the present work is the first clear experimental 
verification of the SDW QCP 
among single-crystalline neutron scattering 
studies on the QCPs of heavy fermions 
and $d$-electron systems. 
Assuming that criticalities of QCPs are classified into 
a limited number of universality classes, we expect that 
the SDW QCP remains to be observed in other systems.

%
In summary, we have demonstrated that the quantum critical 
behavior of the heavy fermion Ce(Ru$_{1-x}$Rh$_{x}$)$_2$Si$_2$ 
is controlled by the SDW type QCP in three space dimensions. 
The inverse correlation time, i.e., energy width 
$\Gamma(\bm{k}_3)$ of the dynamical susceptibility, 
shows a $T^{3/2}$ dependence predicted by 
the renormalization group and SCR theories


\begin{theacknowledgments}
\hspace*{2.5 mm} We wish to acknowledge B. F{\aa}k, J. Flouquet, T. Taniguchi, 
and T. Moriya for valuable discussions.
\end{theacknowledgments}


\bibliographystyle{aipproc}   


\begin{thebibliography}{21}
\expandafter\ifx\csname natexlab\endcsname\relax\def\natexlab#1{#1}\fi
\providecommand{\enquote}[1]{``#1''}
\expandafter\ifx\csname url\endcsname\relax
  \def\url#1{\texttt{#1}}\fi
\expandafter\ifx\csname urlprefix\endcsname\relax\def\urlprefix{URL }\fi
\providecommand{\eprint}[2][]{\url{#2}}

\bibitem[Moriya(1985)]{MoriyaBook85}
T.~Moriya, \emph{Spin Fluctuations in Itinerant Electron Magnetism},
  Springer-Verlag, Berlin, 1985.

\bibitem[{T. Moriya and K. Ueda}(2003)]{Moriya-Ueda03}
{T. Moriya and K. Ueda}, \emph{Rep. Prog. Phys.} \textbf{66}, 1299 (2003).

\bibitem[Hertz(1976)]{Hertz76}
J.~A. Hertz, \emph{Phys. Rev. B} \textbf{14}, 1165 (1976).

\bibitem[Millis(1993)]{Millis93}
A.~J. Millis, \emph{Phys. Rev. B} \textbf{48}, 7183 (1993).

\bibitem[{P. Coleman \textit{et al.}}(2001)]{Coleman01}
{P. Coleman \textit{et al.}}, \emph{J. Phys. Condens. Matter} \textbf{13}, R723
  (2001).

\bibitem[{T. Moriya and T. Takimoto}(1995)]{Moriya-Taki}
{T. Moriya and T. Takimoto}, \emph{J. Phys. Soc. Jpn.} \textbf{64}, 960 (1995).

\bibitem[v.~{L{\"{o}}hneysen \textit{et al.}}(1994)]{Lohneysen94}
H.~v.~{L{\"{o}}hneysen \textit{et al.}}, \emph{Phys. Rev. Lett.} \textbf{72},
  3262 (1994).

\bibitem[{S. Kambe \textit{et al.}}(1996)]{Kambe96}
{S. Kambe \textit{et al.}}, \emph{J. Phys. Soc. Jpn.} \textbf{65}, 3294 (1996).

\bibitem[{S. Raymond \textit{et al.}}(2001)]{raymond01a}
{S. Raymond \textit{et al.}}, \emph{J. Phys.: Condens. Matter} \textbf{13},
  8303 (2001).

\bibitem[Stewart(2001)]{Stewart01}
G.~R. Stewart, \emph{Rev. Mod. Phys.} \textbf{73}, 797 (2001).

\bibitem[{A. Schr{\"{o}}der \textit{et al.}}(2000)]{Schroder00}
{A. Schr{\"{o}}der \textit{et al.}}, \emph{Nature (London)} \textbf{407}, 351
  (2000).

\bibitem[{P. Paschen \textit{et al.}}(2004)]{Paschen04}
{P. Paschen \textit{et al.}}, \emph{Nature (London)} \textbf{432}, 881 (2004).

\bibitem[{Stockert \textit{et al.}}(1998)]{Stockert98}
O.~{Stockert \textit{et al.}}, \emph{Phys. Rev. Lett.} \textbf{80}, 5627
  (1998).

\bibitem[{Q. Si \textit{et al.}}(2001)]{Si01}
{Q. Si \textit{et al.}}, \emph{Nature (London)} \textbf{413}, 804 (2001).

\bibitem[{H. Kadowaki \textit{et al.}}(2005)]{Kadowaki05}
{H. Kadowaki \textit{et al.}}, \emph{cond-mat/0504386}  (2005).

\bibitem[{M. J. Besnus \textit{et al.}}(1985)]{Besnus85}
{M. J. Besnus \textit{et al.}}, \emph{Solid State Commun.} \textbf{55}, 779
  (1985).

\bibitem[Kadowaki et~al.(2004)]{Kadowaki04}
H.~Kadowaki, M.~Sato, and S.~Kawarazaki, \emph{Phys. Rev. Lett.} \textbf{92},
  097204 (2004), and references therein.

\bibitem[{C. Sekine \textit{et al.}}(1992)]{Sekine92}
{C. Sekine \textit{et al.}}, \emph{J. Phys. Soc. Jpn.} \textbf{61}, 4536
  (1992).

\bibitem[{S. Kawarazaki \textit{et al.}}(1997)]{Kawarazaki97}
{S. Kawarazaki \textit{et al.}}, \emph{J. Phys. Soc. Jpn.} \textbf{66}, 2473
  (1997).

\bibitem[{Y. Tabata \textit{et al.}}(1998)]{Tabata98}
{Y. Tabata \textit{et al.}}, \emph{J. Phys. Soc. Jpn.} \textbf{67}, 2484
  (1998).

\bibitem[Sachdev(1999)]{SachdevBook99}
S.~Sachdev, \emph{Quantum Phase Transitions}, Cambridge University Press,
  Cambridge, 1999, {Chap.} 12.

\end{thebibliography}

\end{document}